\documentstyle[a4,12pt]{article}

\newcommand{\NP}[1]{Nucl. Phys.\ {\bf #1}\ }
\newcommand{\PL}[1]{Phys. Lett.\ {\bf #1}\ }

\newcommand{\PR}[1]{Phys. Rev.\ {\bf #1}\ }

\newcommand{\IJMP}[1]{Int. J. Mod. Phys.\ {\bf #1}\ }
\newcommand{\MPL}[1]{Mod. Phys. Lett.\ {\bf #1}\ }
\newcommand{\CQG}[1]{Class. Quantum Grav.\ {\bf #1}\ }
\newcommand{\AP}[1]{Ann. of Phys.\ {\bf #1}\ }

\newcommand{\ba}{\begin{eqnarray}}
\newcommand{\ea}{\end{eqnarray}}
\newcommand{\be}{\begin{equation}}
\newcommand{\ee}{\end{equation}}
\newcommand{\lbl}{\label}

\begin{document}
\begin{titlepage}
\begin{flushright}
CERN-TH.6686/92\\
\end{flushright}
\vspace*{0.5cm}
\begin{center}
{\bf
\begin{Large}
On Anomaly-free Supergravity\\as an Effective String Theory\\
\end{Large}
}
\vspace*{3cm}
         {\large A. K. Tollst\'{e}n}
         \footnote{address after 1 November 1992:
          NORDITA, Blegdamsvej 17,
         DK-2100 Copenhagen \O.}
         \\[.5cm]
         CERN, CH-1211 Geneva 23, Switzerland
\end{center}
\vspace*{2cm}
\begin{abstract}
The equations of motion of  anomaly-free supergravity are shown to
fulfil (to all orders in $\alpha'$)
a  differential condition corresponding to
the one relating the Weyl anomaly
coefficients for a non-linear sigma model representing a (heterotic)
string propagating in a non-trivial background. This supports  the
possibility that anomaly-free supergravity could provide the complete
massless
effective theory for the heterotic string.
\end{abstract}
\vfill
CERN-TH.6686/92\\
October 1992
\end{titlepage}
When studying the propagation of strings in background fields, the
conditions for conformal invariance of the non-linear
sigma model
coincide, as is well known, to the lowest order with the
equations of motion for 10-dimensional supergravity, coupled to
Yang-Mills in the case of the heterotic string. See, for
instance, \cite{CT} for an introductory review.
Imposing conformal
invariance also at quantum level thus gives ``stringy" corrections
to these equations, with $\alpha'$ the sigma-model loop expansion
parameter.
The dilaton equation
turns out to play the role of the central charge of the corresponding
Virasoro algebra \cite{CFMP85,dA86},
and must thus be independent of position in spacetime.
This is indeed true since it has
been shown that the Weyl anomaly coefficients satisfy a ``Bianchi
identity" of the form
\be
\lbl{1}
D^\mu\beta{\mu\nu}{}^{(g)}=\partial_\nu \beta^{(\phi)}
\ee
once the other equation(s) are imposed. Hence, if the l.h.s. vanishes,
so does the
derivative of the dilaton equation, $\beta^{(\phi)}$, and it must then
itself be a constant. This identity was later proven to all orders in
$\alpha'$ for the bosonic sigma model \cite{CP86,T87,O89},
and a supersymmetric version of it was made in \cite{H89}.

A given sigma model corresponds to a string theory only if there exists
a Virasoro algebra, which requires a condition like (\ref{1}).
Another way to regard this equation is to observe that it is a direct
consequence of D-dimensional general covariance of the effective action
for the supergravity theory \cite{dA86,T87}.
We can now turn the argument around: Given a set of dynamical equations
for a (super)gravity theory, are they derivable from an action,
and is this also the effective theory corresponding to a string
propagating in a non-trivial background?
As long as we do not succeed in deriving the background action
by integrating out the string coordinates, this question can only
be answered by an explicit calculation, order by order
in $\alpha'$. However, the Bianchi identity (\ref{1}) provides us with a
necessary condition that our candidate must satisfy.

The so-called anomaly-free supergravity (AFS) is a model of supergravity
coupled to a Yang-Mills theory in ten dimensions containing (implicit)
corrections to all orders in $\alpha'$.  It is obtained by imposing
constraints on the superspace Bianchi identities,
with the Lorentz Chern-Simons term added in
the way required for gauge and gravitational
anomaly cancellations \cite{GS84}, and then solving for the
physical fields. Thanks to the observation made by Bonora, Pasti and
Tonin \cite{BPT87}
that the new $O(\alpha')$ terms do not add any new irreducible
representations of $SO(1,9)$ to the equations, these can indeed be
solved explicitly, albeit after very cumbersome calculations
\cite{BBLPT88,DaFRR88,P91}. The
solution consists of a set of equations of motion, supersymmetry
transformations, and $x$-space Bianchi identities. They contain, as
already mentioned, implicit corrections to all orders in $\alpha'$.
To express these equations in physical fields only, we need to
solve the equation relating   $H_{\mu\nu\rho}$ to the torsion order by
order, and we thus obtain corrections to all orders. The lowest order(s)
coincide (after field redefinitions) with the effective theory for the
heterotic string.

Unfortunately, AFS is not unique, so it need not be {\em the} heterotic
string effective theory. The relation between the torsion and the
physical fields is a differential equation, and different theories
might emerge as a result of different boundary conditions.
It is also, at least in principle,  possible to add
non-minimal terms by keeping more $SO(1,9)$ representations non-zero in
the constraints imposed. Both these mechanisms have been suggested
\cite{DaFRR88,LP(T)} to
account for the $\zeta (3) R^4$ corrections \cite{zeta3},
and string loop  terms should somehow turn up this way too.
Strong restrictions on possible non-minimal terms must be provided
by the fact that
AFS and the heterotic sigma model have in common
a non-trivial instanton solution
\cite{PT91}.

The  purpose of this letter is to strengthen
the case that AFS is indeed closely related to
the full effective  theory by  showing  that
its bosonic equations of motion satisfy an equation corresponding to
(\ref{1}) to all orders in $\alpha'$.

In  \cite{P91}  the
bosonic parts of the equations of motion of AFS are given explicitly.
We get the dilaton equation

\be
\lbl{10}
\beta \equiv \Box\phi + \frac{1}{6}T^2 + \frac{4}{3}(\partial\phi)^2
- 2 e^{-\frac{4}{3}\phi}\mbox{tr}F^2
-\gamma_1 e^{-\frac{4}{3}\phi} W = 0 ,
\ee
and Einstein's equations can be written
\ba
\lbl{11}
\beta_{\mu\nu} & \equiv & R_{\mu\nu}
+\frac{2}{3}D_{(\mu}\partial_{\nu)}\phi +
\frac{8}{9}\partial_\mu\phi\partial_\nu\phi+g_{\mu\nu}
\left(\frac{1}{3}\Box\phi+\frac{4}{9}(\partial\phi)^2\right)\nonumber
\\
& - & 8e^{-\frac{4}{3}\phi}\mbox{tr}(F_{\mu\rho} F_\nu{}^\rho )
- \gamma_1
e^{-\frac{4}{3}\phi}\left[W_{\mu\nu}+
\frac{1}{2}\left(\frac{2}{27}+h_1\right)g_{\mu\nu}\Box
T^2\right] = 0.
\ea
Here
\ba
\lbl{12}
W &=& -R_{\mu\nu}R^{\mu\nu}+\frac{1}{2}R_{\mu\nu\rho\sigma}
R^{\mu\nu\rho\sigma}  \nonumber   \\
&+& \frac{1}{2}\left(-\frac{1}{9}+3h_1\right)\Box T^2 -
\frac{2}{3}D_{[\mu}T_{\nu_1.\nu_3]}D^\mu
T^{\nu_1.\nu_3}-4T^{\mu_1\mu_2\rho}T^{\nu_1\nu_2}{}_\rho
D_{\mu_1}T_{\mu_2\nu_1\nu_2} \nonumber \\
&-&2(T_\mu T_\nu)(T^\mu T^\nu) + 4 (T_\mu T_\nu
T^\mu T^\nu) + \frac{1}{3}\left(\frac{2}{27}+h_1\right)(T^2)^2 ,
\ea
and
\ba
\lbl{13}
W_{\mu\nu} &=& -\Box R_{\mu\nu} - 4R_{\mu\rho}R_\nu{}^\rho +
2R_{\rho_1\rho_2\sigma\mu} R^{\rho_1\rho_2\sigma}{}_\nu \nonumber \\
&+& \left(\frac{2}{27}+h_1\right)D_{(\mu}\partial_{\nu)}T^2
-\frac{8}{3}g_{\nu\sigma}D_{[\mu}T_{\rho_1.\rho_3]}D^{[\sigma}
T^{\rho_1.\rho_3]}
+ 2\left(\frac{2}{27}+h_1\right)R_{\mu\nu}T^2 \nonumber \\
&-& 8T_\mu{}^{\rho_1\sigma}T_\sigma{}^{\rho_2\rho_3}
D_{[\nu}T_{\rho_1.\rho_3]}
- 8T_\nu{}^{\rho_1\sigma}T_\sigma{}^{\rho_2\rho_3}
D_{[\mu}T_{\rho_1.\rho_3]} \nonumber \\
&+& 8 T_{\mu\rho\tau}T_{\nu\sigma}{}^\tau(T^\rho T^\sigma) +
16(T_\mu T_\rho T_\nu T^\rho ).
\ea
The curvature tensor is defined as
\be
\lbl{131}
R_{\mu\nu\rho}{}^\sigma = \partial_{[\mu}\tilde{\Gamma}_{\nu ]\rho}
{}^\sigma -\tilde{\Gamma}_{[\mu |\rho |}{}^\tau
\tilde{\Gamma}_{\nu ]\tau}{}^\sigma  ,
\ee
where $\tilde{\Gamma} = \Gamma +T $, and the torsion is
\be
\lbl{132}
T_{\mu\nu\rho} = -3 H_{\mu\nu\rho}e^{-\frac{4}{3}\phi}
-2\gamma_1 e^{-\frac{4}{3}\phi} W_{\mu\nu\rho}   ,
\ee
with
\be
\lbl{20}
W_{\mu\nu\rho} = \frac{1}{2}\Box T_{\mu\nu\rho} +
3T_{[\mu}{}^{\sigma_1\sigma_2}R_{\nu\rho ]\sigma_1\sigma_2}
+3T_{[\mu\nu}{}^\sigma R_{\rho ]\sigma} -4(T_{[\mu} T_\nu T_{\rho
]})-\left(\frac{2}{27}+h_1\right)T^2T_{\mu\nu\rho} .
\ee

The use of the notation ``$\beta$" and ``$\beta_{\mu\nu}$"
does not imply that the above expressions are $\beta$-functions or
Weyl anomaly coefficients. Since AFS only exists as an on-shell
theory, all we know is that they are linear combinations of
the equations of motion, as we will
see below, when we study the lowest-order action.
We employ the
shorthand notation $(T_\mu T_\nu )= T_{\mu\rho_1}{}^{\rho_2}
T_{\nu\rho_2}{}^{\rho_1}$ and similarly for the higher ``traces"
and $ T^2 =
T_{\mu_1.\mu_3}T^{\mu_1.\mu_3}$. The coefficient $\gamma_1$ introduced
in \cite{DaFRR88}
is proportional to $\alpha'$ once AFS is interpreted as an
effective string theory, and $h_1$ is an arbitrary
constant, which can be removed by a redefinition of $\phi$ in terms
of $T^2$.
We also have the trace of (\ref{11})
\be
\lbl{201}
\beta_\mu{}^\mu = R-\frac{2}{3}T^2  ,
\ee
which we regard as a constraint, and
the remaining equations of motion are
\be
\lbl{14}
D^\mu F_{\mu\nu}=0
\ee
\be
\lbl{141}
D^\mu T_{\mu\nu\rho}=0.
\ee
The AFS version of (\ref{1}) can rather easily be
found by trial and error, but a nicer way is to use \cite{dA86} for the
action with $\gamma_1=0$. This action can be obtained from equation
(10.4) of \cite{DaFRR88}, and is
\be
\lbl{142}
S=\int d^{10} x \sqrt{g} e^{\frac{4}{3}\phi}\left( R +
\frac{3}{2}e^{-\frac{8}{3}\phi}H^2 - 4e^{-\frac{4}{3}\phi}\mbox{tr}F^2
\right) .
\ee
Here we use an $R$ depending only on the metric, and have expressed
$T$ in $H$.
We then find that (the zero superscript of course denotes $\gamma_1=0$)
\be
\lbl{143}
\frac{e^{-\frac{4}{3}\phi}}{\sqrt{g}}\frac{\delta S}{\delta g^{\mu\nu}}
= \beta_{\mu\nu}^{(0)}-\frac{g_{\mu\nu}}{2}(\beta_\rho^{(0)}{}^\rho
+ 2\beta^{(0)} )
\ee
and
\be
\lbl{144}
\frac{e^{-\frac{4}{3}\phi}}{\sqrt{g}}\frac{\delta S}{\delta \phi}
= \frac{4}{3}\beta_\rho^{(0)}{}^\rho .
\ee
The ``Bianchi identity" can be written
\be
\lbl{145}
D^\mu \frac{\delta S}{\delta g^{\mu\nu}} + \frac{1}{2}\partial_\nu \phi
\frac{\delta S}{\delta \phi} = 0    ,
\ee
which is exactly the condition for invariance of the action under
$\delta x^\mu= V^\mu$,  $V^\mu$ being an
arbitrary vector field,
and if we insert (\ref{143}) and (\ref{144}), directly imposing the
constraint $\beta_\rho^{(0)}{}^\rho = 0$, we find
\be
\lbl{146}
D^\mu\left(e^{\frac{4}{3}\phi}\beta_{\mu\nu}^{(0)}\right)-\partial_\nu
\left(e^{\frac{4}{3}\phi}\beta^{(0)}\right) = 0   .
\ee

We will now try to prove this equation extended to all orders in
$\gamma_1$.
To do this we will need various identities.
{}From (\ref{131}) we get the symmetries of the curvature
\ba
\lbl{15}
R_{\mu\nu\rho\sigma}-R_{\rho\sigma\mu\nu} &=& D_{[\mu}T_{\nu ]\rho\sigma}
 - D_{[\rho}T_{\sigma ]\mu\nu}\\
R_{\mu [\nu_1 . \nu_3]} &=& \frac{1}{2}\left( D_\mu T_{\nu_1 . \nu_3}
-D_{[\nu_1}T_{\nu_2 \nu_3]\mu}\right)
+2T_{\mu [\nu_1}{}^\rho T_{\nu_2\nu_3 ]
\rho} \\
R_{[\nu_1.\nu_3]\mu} &=& D_{[\nu_1}T_{\nu_2\nu_3 ]\mu}
+2T_{[\nu_1\nu_2}{}^\rho T_{\nu_3 ]\mu \rho}
\ea
the Bianchi identity
\be
\lbl{16}
D_{[\mu}R_{\nu\rho ] \sigma}{}^\tau = 2T_{[\mu\nu}{}^\lambda R_{\rho ]
\lambda\sigma}{}^\tau
\ee
and its contractions
\be
\lbl{17}
2D_{[\mu}R_{\nu ]\rho}+D_\sigma R_{\mu\nu\rho}{}^\sigma =
-2T_{\mu\nu}{}^\sigma R_{\rho\sigma}+
4 T_{\sigma [ \mu}{}^\tau R_{\nu ] \tau\rho}{}^\sigma \\
\ee
\be
\lbl{171}
D^\mu R_{\mu\nu} = \frac{1}{12}\partial_\nu T^2 + \frac{1}{2}T^{\mu_1
.\mu_3}D_{\mu_1}T_{\mu_2\mu_3\nu}  .
\ee
Furthermore, for the gauge field and the torsion, the Bianchi
identities are
\be
\lbl{18}
D_{[\mu}F_{\nu\rho ]}= 2 T_{[\mu\nu}{}^\sigma F_{\rho ]\sigma}
\ee
\ba
\lbl{19}
D_{[\mu}T_{\nu\rho\sigma]} &=& -\frac{4}{3}\partial_{[\mu}\phi
T_{\nu\rho\sigma]} - 3T_{[\mu\nu}{}^\tau T_{\rho\sigma ]\tau}+ 12
e^{-\frac{4}{3}\phi}\mbox{tr}(F_{[\mu\nu}F_{\rho\sigma]}) \nonumber \\
&+& \gamma_1e^{-\frac{4}{3}\phi}\left[-2D_{[\mu}W_{\nu\rho\sigma]}-
6T_{[\mu\nu}{}^\tau W_{\rho\sigma ]\tau}
+3R_{[\mu\nu}{}^{\tau\lambda}R_{\rho\sigma ] \tau \lambda}\right].
\ea
Equation (\ref{19}) is obtained
by combining the Bianchi identity for $H$
{}from \cite{P91} with (\ref{132}).
We also have the Ricci identity (for an arbitrary covariant vector,
$V_\rho$)
\be
\lbl{21}
[D_\mu,D_\nu]V_\rho = -2R_{\mu\nu\rho}{}^\sigma V_\sigma
-2T_{\mu\nu}{}^\sigma D_\sigma V_\rho  ,
\ee
{}from which we immediately derive the useful commutator
\be
\lbl{22}
[D_\mu,\Box ] V_\nu = -2\left[ D^\rho R_{\mu\rho\nu}{}^\sigma V_\sigma +
2R_{\mu\rho\nu}{}^\sigma D^\rho V_\sigma
- R_\mu{}^\rho D_\rho V_\nu \right]   .
\ee

After a lenghty calculation, the details of which are of little interest,
we can now prove
\be
\lbl{30}
D^\mu\left(e^{\frac{4}{3}\phi}\beta_{\mu\nu}\right)-
\partial_\nu\left(e^{\frac{4}{3}\phi}\beta\right)=0  .
\ee
The lowest order in $\gamma_1$ is easy. Using our identities and
the equations of motion for $F$ and $T$, everything vanishes except a
higher-order contribution from (\ref{19}). We are then left with the
$O(\gamma_1)$ part
\ba
\lbl{31}
\lefteqn{
D^\mu W_{\mu\nu} +\frac{1}{2}\left(\frac{2}{27} +
h_1\right)\partial_\nu\Box T^2-\partial_\nu W}\nonumber \\
& & -\frac{4}{3}T^{\mu_1.\mu_3}D_{[\nu}W_{\mu_1.\mu_3]}
-2T^{\mu_1.\mu_3}T_{\nu\mu_1}{}^\rho W_{\mu_2\mu_3\rho} +2
T^{\mu_1.\mu_3}R_{\nu\mu_1}{}^{\rho\sigma}R_{\mu_2\mu_3\rho\sigma} .
\ea
Written out explicitly, this is a very long expression containing terms
of the form $D\Box R$, $RDR$, $R^2T$, $TD^3 T$, $DTD^2T$, $DR T^2$,
$RTDT$, $T^2D^2T$, $RT^3$, $T(DT)^2$, and $T^3DT$. ($T^5$ terms with one
free index cannot exist for symmetry reasons.) A systematic elimination
using (\ref{201}),
(\ref{141}), (\ref{15}) -- (\ref{171}), and the Ricci identity
leaves us with only terms of the two last types,
which separately cancel.

We have thus proven that the equations of motion derived in the AFS
scheme do indeed fulfil the differential condition to all orders in
$\alpha'$. The crucial point is that we need not use (\ref{19})
in the $O(\gamma_1)$ part of the calculation, so we never need to go
higher than the first order in $\gamma_1$ explicitly. Hence AFS
satisfies a necessary condition for the existence of an action from
which its equations of motion can be derived, and of a string theory,
the Weyl anomaly conditions of which are the equations of motion of AFS.
This might also give some useful hints for the derivation of the full
AFS Lagrangian (cf. \cite{CKP86,T287}).
However, the fact that $H$ and $T$
 are related
via a differential constraint might still make this a very difficult
task. If the full AFS theory  including the fermions is also
derivable from an action, (\ref{30}) should still hold, and the full
theory might then be an effective theory for the heterotic string
with a non-trivial {\em fermionic} background. It would be rather
interesting to have this, since very little work has been done with such
sigma models; see however \cite{CFMP85,FMS}.

The question is now which parts of string theory can be accounted
for by our effective theory.
We do not know whether
string loop effects are also contained in AFS, but we can argue
that they are  unlikely to occur in the minimal model:
To obtain the sigma model action we have
to make the rescaling $g_{\mu\nu}\rightarrow e^{-\frac{4}{3}\phi}
g_{\mu\nu}$, remembering that the ``fundamental" torsion is
the one with two covariant and one contravariant indices, so that
$T_{\mu\nu\rho}$ has to be rescaled as $g_{\mu\nu}$.
The lowest-order
action then has the form $S=\int d^{10}x \sqrt{g} e^{-4\phi}{\cal L}$,
where $\cal L$ does not contain any exponentials, and these can also
be divided out from the equations of motion and the relation
between $H$ and $T$. Adding higher-order terms and
solving for $H$ should then introduce
no terms with a different power of $e^\phi$, that is the
string coupling constant, anywhere. Nevertheless, unless
supersymmetry is explicitly broken by string loops there must
exist a set of constraints, perhaps very complicated, extending
minimal AFS to a model also containing these effects.

Could minimal AFS then be the full effective theory corresponding
to string tree level? We already know that there is trouble
with the $\zeta(3) R^4$ term, since no transcendental coefficients
turn up naturally. As was mentioned already in the introduction
two possible scenarios have been proposed; it is an effect of
the choice of boundary conditions for the solution of (\ref{132}),
or it is non-minimal \cite{DaFRR88,LP(T)}. An argument in favour
of the latter suggestion is that this term can be separately
supersymmetrized, at least if an (off-shell) superfield exists
\cite{R4}. However, in a very recent paper by de Roo et al. \cite{RSW92}
it is argued that the supersymmetrization  clashes
with gauge-invariance if only physical fields are present, while
there is no such problem with the similar $R^4$ term coming from
string loops \cite{loops}.
It is then
very tempting to believe that what we have is the full effective
theory corresponding to the tree level heterotic string, the
apparent contradiction found in \cite{RSW92} perhaps being resolved
by the addition of higher-order corrections to the supersymmetry
transformations, and that loop effects would be accounted for
by the addition of non-minimal terms.
This should clearly be examined by  calculating the
corrections predicted by AFS  to cubic order in $\alpha'$ and
comparing to known results, see for instance \cite{RSW92,BR89,FMR88}
and references therein. The reason why this has not already been
done is probably the difficulty expected in handling (\ref{132}).
In view of the rather  strong indications that AFS has to
be taken seriously as an effective string theory,
compactifications and other classical solutions
should also be studied via AFS. This could  be an easier
problem, since it might be unnecessary to solve for $H$ explicitly.
Some attempts in this direction
have already been made in the second reference of \cite{P91} and  in
\cite{PT91}.

\end{document}